\documentstyle[prl,aps,psfig]{revtex}

\newcommand{\bra}[1]{\langle #1 | \,}
\newcommand{\ket}[1]{\, | #1 \rangle}
\newcommand{\om}{\omega}
\newcommand{\ga}{\gamma}
\newcommand{\eps}{\epsilon}
\newcommand{\De}{\Delta}

\begin{document}

\draft


\title{Phase-Control of Photoabsorption in Optically Dense Media}

\author{David Petrosyan$^{1,2}$ and P.~Lambropoulos$^{1,3,4}$}

\address{$^1$ Institute of Electronic Structure \& Laser, 
FORTH, P.O. Box 1527, \\
Heraklion 71110, Crete, Greece \\
$^2$ Institute for Physical Research, ANAS,
Ashtarak-2, 378410, Armenia \\
$^3$ Max--Planck--Institut f\"ur Quantenoptik,
Hans--Kopfermann--Stra{\ss}e 1, \\
D--85748 Garching, Germany \\
$^4$ Department of Physics, University of Crete, Greece}

\date{\today}

\maketitle

\begin{abstract}

We present a self-consistent theory, as well as an illustrative application 
to a realistic system, of phase control of photoabsorption in an optically 
dense medium.  We demonstrate that, when propagation effects are taken 
into consideration, the impact on phase control is significant.  
Independently of the value of the initial phase difference between the two 
fields, over a short scaled distance of propagation, the medium tends to 
settle the relative phase so that it cancels the atomic excitation. 
In addition, we find some rather unusual behavior for an optically thin layer.

\end{abstract}



\noindent
After the initial ideas \cite{shapiro} and experimental demonstration 
\cite{elliott,zhu} of the feasibility of the control of photoabsorption 
and its products through the control of the relative phase of two fields, 
much work in atoms \cite{nl} and molecules \cite{zhu} has explored a variety 
of processes.  Many interesting issues \cite{zhu,nl,cl} have been raised 
and clarified, establishing thus the idea as a useful tool. With the exception
of one paper \cite{elliott_pr}, however, theory and experiment have dealt only
with single-atom (molecule) situations.  But, if the ideas are to be 
contemplated for applications, the issue of propagation is crucial.  
Addressing this issue is the purpose of this Letter.

We have chosen the fundamental scheme \cite{elliott,cl} which has served as a 
benchmark for many of the initial and continuing work. We consider the 
excitation of a bound transition through the combined effect of a single- 
and a three-photon transition via two fields whose relative phase is 
controlled externally. We formulate and examine the propagation of a 
bichromatic electromagnetic field $E$ through an optically dense medium 
consisting of Xe atoms. This electric field is a function of time $t$ and 
space coordinate $z$ and is composed of the fundamental and its third harmonic
fields that have the same (linear) polarization and frequencies $\om_f$ and 
$\om_h=3\om_f$, respectively. It is expressed as
\begin{equation}
E(z,t) = \frac{1}{2} [E_f e^{i(k_f z - \om_f t)}	+ 
E_h e^{i(k_h z - \om_h t)} + \text{c. c.}] , \label{E_total}
\end{equation}
where $E_j={\cal E}_j e^{-i \phi_j}$, $j=f,h$, with ${\cal E}_j$ and
$\phi_j $ the slowly varying in time and space real amplitude and phase
of the corresponding field, and $k_j= \om_j n_j c^{-1}$, with $n_j$ the 
refraction index of the host medium at frequency $\om_j$. Although in 
our present treatment the host medium is vacuum, and thus $n_f=n_h=1$ and 
$k_h=3k_f$, for the sake of generality, e.g., presence of a buffer gas, we 
shall keep in the formalism the refraction index. The frequencies 
$\om_{h,f}$ are chosen so that one harmonic photon and three fundamental 
photons are at near resonance with the transition from the ground state 
($\ket{1}$) to the $6s$ state ($\ket{2}$) of Xe. A two-photon transition 
due to the strong fundamental or one-photon transition due to the harmonic 
fields lead to the ionization continuum (states $\ket{c}$) of the atom. 
As we intend to explore intensities of the fields for which the
one- and three-photon transition amplitudes between states $\ket{1}$ and 
$\ket{2}$ are of comparable magnitude so as to maximize the modulation depth, 
the transition $\ket{2} \to \ket{c}$ would be dominated by the two-photon 
process and the one-photon ionization due to harmonic photon can be neglected.
Experimental contexts for the situation we are considering have been detailed
in the literature. 

Beginning with the second order wave equation for the field $E(z,t)$, 
in the slowly varying (during an optical cycle) amplitude approximation 
one neglects  all second derivatives and, after projecting onto the 
corresponding mode function $\exp[i(\om_j t - k_j z)]$, $j=f,h$, one arrives at
\begin{equation}
\frac{\partial E_j}{\partial z} + 
\frac{n_j}{c} \frac{\partial E_j}{\partial t} = 
\frac{1}{c \eps_0 n_j} 
\left[ i \frac{\om_j}{2}P_j - \frac{\partial P_j}{\partial t} \right] , 
\label{maxw}
\end{equation}
where $P_j=P_j^{\prime} e^{-i \phi_j}$ ($P_j^{\prime}$ being complex) is the 
slowly varying in time and space field-induced medium polarization at frequency
$\om_j$. Consistently with Eq. (\ref{E_total}), it can be expressed as
\begin{equation}
P(z,t) = \frac{1}{2} [P_f e^{i(k_f z - \om_f t)}	+ 
P_h e^{i(k_h z - \om_h t)} + \text{c. c.}] . \label{P_totalF}
\end{equation}
The most general approach to the calculation of the response of the medium 
is through the atomic density matrix $\rho$ which obeys the equation 
$ \partial_t \rho = -i \hbar^{-1}[H_{\text{atom}} + D ,\rho ]$, with 
$H_{\text{atom}}$ the free atomic Hamiltonian and $D = - \mu E$ the 
atom-field interaction in the dipole approximation where $\mu$ is the 
electric dipole operator. Introducing the rotating wave approximation 
and adiabatically eliminating the continuum and all virtual (nonresonant)
bound states (connecting by the lowest-order paths the states $\ket{1}$ and
$\ket{2}$), the slowly varying density matrix elements of the two remaining 
states $\sigma_{11}\simeq \rho_{11}$, $\sigma_{22}\simeq \rho_{22}$, and 
$\sigma_{21}\simeq \rho_{21}\exp[i 3 ( \om_f t + \phi_f - k_f z )]$, are
found to obey the following set of equations:   
\begin{mathletters}
\label{sigs}
\begin{eqnarray}
\frac{\partial }{\partial t} \sigma_{11} &=& \ga \sigma_{22} - 
\text{Im} \left[ \left( \frac{\mu_{12}^{(3)}}{\hbar} {\cal E}_f^3 + 
e^{i\theta}\frac{\mu_{12}}{\hbar} {\cal E}_h  
\right) \sigma_{21} \right] ,  \label{sig11} \\
\frac{\partial }{\partial t} \sigma_{22} &=& 
- (\ga+\ga_{\text{ion}}) \sigma_{22} + 
\text{Im} \left[ \left( \frac{\mu_{12}^{(3)}}{\hbar} {\cal E}_f^3 + 
e^{i\theta}\frac{\mu_{12}}{\hbar} {\cal E}_h  
\right) \sigma_{21} \right] ,  \label{sig22} \\
\frac{\partial }{\partial t} \sigma_{21} &=& 
- \left[ \frac{\ga+\ga_{\text{ion}}}{2} +
i \left(\De - 3 \frac{\partial \phi_f}{\partial t}  \right) + 
i \frac{s_1-s_2}{2\hbar} {\cal E}_f^2 \right] \sigma_{21}   
\nonumber \\ & & +
i \left( \frac{\mu_{12}^{(3)}}{2\hbar} {\cal E}_f^3 + 
e^{-i\theta}\frac{\mu_{12}}{2\hbar} {\cal E}_h  
\right) (\sigma_{11} - \sigma_{22} )   ,  \label{sig21} 
\end{eqnarray}
\end{mathletters}
where $\ga $ is the radiative decay rate of level $\ket{2}$, 
$\ga_{\text{ion}} \propto (\mu_{2c}^{(2)} I_f)^2$ is the 2-photon ionization
rate of $\ket{2}$ being proportional to the square of the intensity 
$I_f \propto {\cal E}_f^2$ of the fundamental ($\mu_{2c}^{(2)}$ is the 
effective 2-photon matrix element for the fundamental field on the transition 
$\ket{2} \to \ket{c}$), $\mu_{12}^{(3)}$ is the effective 3-photon matrix 
element for the fundamental field on the transition $\ket{1} \to \ket{2}$, 
$s_1$ and $s_2$ are the lowest-order Stark shift coefficients 
(polarizabilities) of levels $\ket{1}$ and $\ket{2}$, respectively, and
$\mu_{12} \equiv \bra{1} \mu \ket{2}$ is the matrix element of the electric 
dipole operator $\mu$. Finally $\De$ is the detuning of both fields from the 
$\ket{1} \to \ket{2}$ transition resonance and 
$\theta = (\phi_h-3\phi_f)-(k_h-3k_f)z$ their relative phase. 

Consider now the polarization $P(z,t) = N \text{Tr}[\mu \rho]$ of a medium 
of atomic density $N$. In expanding the trace of this equation, we again 
follow the same procedure as in obtaining Eqs. (\ref{sigs}), i.e. we use 
the adiabatic approximation to expresses all density matrix elements 
that do not refer to the states $\ket{1}$ and $\ket{2}$ in terms of the three 
main elements $\sigma_{11}$, $\sigma_{22}$, and $\sigma_{21}$. Equating the 
result with Eq. (\ref{P_totalF}), identifying and grouping together terms 
oscillating with the same frequencies, we obtain 
\begin{mathletters}
\label{pols}
\begin{eqnarray}
P_f^{\prime} &=& 2 N [{\cal E}_f (s_1 \sigma_{11} + s_2 \sigma_{22}) +
3 \mu_{12}^{(3)}{\cal E}_f^2 \sigma_{21} + 
i \pi \hbar^{-1} |\mu_{2c}^{(2)}|^2 {\cal E}_f^3 \sigma_{22} ] , 
\label{pol_f} \\  
P_h^{\prime} &=& 2 N \mu_{12} \sigma_{21} e^{i \theta} . \label{pol_h}
\end{eqnarray}
\end{mathletters}
Those equations, together with the Maxwell's Eq. (\ref{maxw}) and the atomic
density matrix Eqs. (\ref{sigs}), provide a complete description of our system
in terms of a closed set of equations.

To present the numerical results for Xe, we use the parameters calculated 
previously \cite{cl} via MQDT and appropriately converted to conform
to the present definitions. For illustration purposes, it is desirable to 
have a maximally pronounced interference of the fundamental and harmonic 
fields. The respective Rabi frequencies are given by the first and second 
terms in the parentheses of Eq. (\ref{sig11}). To obtain, for example complete
cancellation at $\theta = \pi$ when these two terms are purely real and have 
opposite signs, it is obvious that the peak values and the temporal widths of 
both Rabi frequencies should be equal so as to overlap completely. Let the 
strong fundamental field have a Gaussian temporal profile with a peak 
amplitude ${\cal E}_f^{\text{max}} \equiv {\cal E}_f (t=t_{\text{max}})$ and 
width $\tau_f$. Then the peak amplitude and width of the weak harmonic field 
should satisfy the relations 
\begin{equation}
{\cal E}_h^{\text{max}}= \frac{\mu_{12}^{(3)}}{\mu_{12}} 
({\cal E}_f^{\text{max}})^3  , \; \;
\tau_h = \frac{\tau_f}{\sqrt{3}} . \label{ampl_wid}
\end{equation} 

\begin{figure}[htb]
\centerline{\psfig{figure=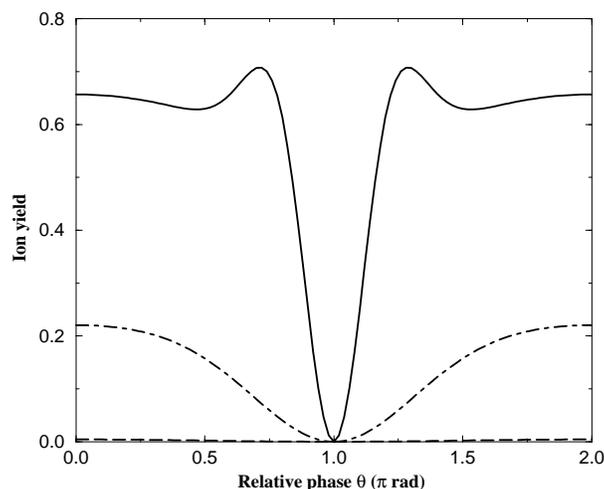,width=8cm}}
\caption{Ion yield 
$Q = (1-\sigma_{11}-\sigma_{22})_{t \to \infty} 
\simeq 1-\sigma_{11}(t \to \infty)$ 
versus relative phase $\theta$ for three different peak intensities of the 
fundamental: 
$I_f^{\text{max}} = 1\times 10^{10}$ W/cm$^2$ (dashed line),
$I_f^{\text{max}} = 3\times 10^{10}$ W/cm$^2$ (dot-dashed line),
$I_f^{\text{max}} = 8\times 10^{10}$ W/cm$^2$ (solid line).}
\label{ion_phase}
\end{figure}

In Fig. \ref{ion_phase} we plot the ion yield 
$Q=[1-\sigma_{11}(t)-\sigma_{22}(t)]_{t \to \infty}$ at 
$z = 0$ as a function of the relative phase $\theta$ for three different 
intensities $I_f$ of the fundamental. In all cases, the detuning $\De$ is 
taken such that it compensates the relative Stark shift of levels $\ket{1}$ 
and $\ket{2}$ at the maximum $t_{\text{max}}$ of the pulse, the harmonic pulse
duration $\tau_h = 1$ ns, and the conditions (\ref{ampl_wid}) are satisfied. 
In this figure, for all intensities and relative phase $\theta = \pi$, the 
ionization vanishes completely since the two transition amplitudes interfere 
destructively and the second term on the rhs of Eq. (\ref{sig11}), responsible
for the stimulated transition from $\ket{1}$ to $\ket{2}$ is equal to zero 
throughout the duration of the pulses. Consequently, the medium practically 
does not interact with the fields and the atoms are ``trapped'' in their 
ground state $\ket{1}$. The more surprising result, 
however, is that in the case $I_f^{\text{max}} = 8\times 10^{10}$ W/cm$^2$, 
maximal ionization is found not for $\theta = 0, 2\pi$, as one 
would expect and is the case for the other intensities. This is a 
manifestation of the quantum-mechanical interference resulting from the 
fact that for this set of parameters, in Eqs. (\ref{sigs}) the terms 
responsible for the stimulated transition reach the maxima at 
$\theta \simeq \pi \pm 0.28 \pi$ where the ionization peaks are located. 
The numerical simulations also show that, while keeping the conditions 
(\ref{ampl_wid}) satisfied, with decreasing pulse duration $\tau_f$, 
the ion yield reduces and its peaks at $\theta \neq 0,2\pi ,...$ gradually 
disappear, which is analogous to the decreasing of intensity since the total 
energy of the pulse lessens. Increasing the intensity, however, results in 
a narrower dip in the ionization profile and a shift of its peaks towards 
the values of $\theta$ that are closer to $\pi$.  

Let us turn now to the propagation effects. The results presented below are 
obtained for a density of atoms $N=10^{13}$ cm$^{-3}$. This, however, does 
not imply any limitation on the generality of the discussion since, as one 
can easily verify, the parameter $z N \Sigma$, where $\Sigma$ is the laser 
beam cross-section, is a propagation constant, and thus it is always possible 
to rescale the problem to any desired density and propagation length $z$. 
Conditions (\ref{ampl_wid}) are assumed at the entrance to the medium. 
As we have noted above, in the case of initial phase difference  
$\theta (0,t) = \pi$, the atoms stay in the ground state and the medium 
appears to be ``transparent'' to both fields; neither the fundamental, nor 
the harmonic experience any remarkable distortion of their shapes or total 
energy $S_j (z) \propto \int dt |{\cal E}_j(z,t)|^2$, $j=f,h$, over distances 
of propagation $z$ as large as $ \sim 50$ cm. The accumulated over this 
distance change of the relative phase is only $\sim 10^{-3} \pi$ rad, which 
is due to the field independent phase shift of the fundamental, given by the 
term in parentheses of Eq. (\ref{pol_f}).

\begin{figure}[htb]
\centerline{\psfig{figure=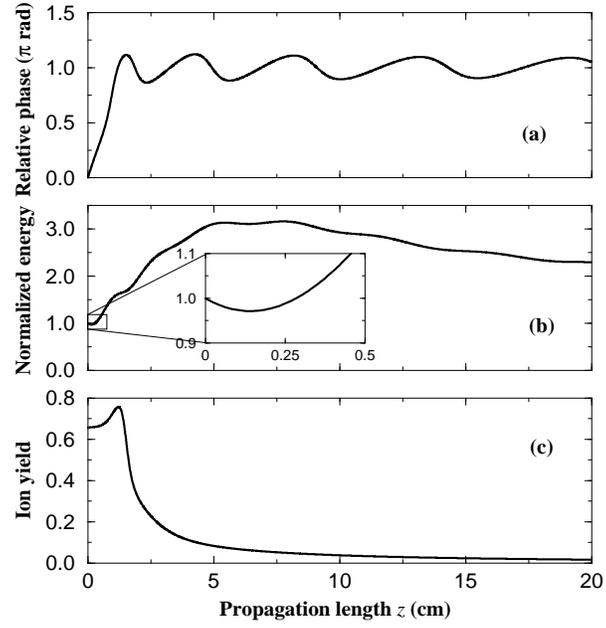,width=8cm}}
\caption{Relative phase $\theta(z,t=t_{\text{max}}+z/c)$ (a), 
normalized energy $S_h(z)/S_h(0)$ of harmonic field (b), and  
ion yield $Q(z)$ (c) versus propagation length $z$ for the case 
$I_f^{\text{max}} = 8\times 10^{10}$ W/cm$^2$.}
\label{ioenph_z}
\end{figure}

\begin{figure}[htb]
\centerline{\psfig{figure=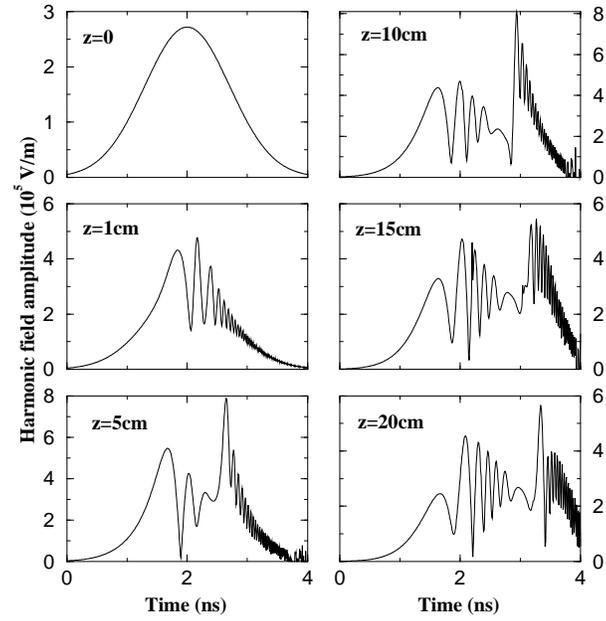,width=8cm}}
\caption{Temporal profile of the amplitude ${\cal E}_h$ of harmonic field 
at different $z$. All parameters are as in Fig. \protect\ref{ioenph_z}.}
\label{pulse_h}
\end{figure}

Consider next the case $\theta(0,t)=0$, i.e., at the entrance to the cell 
the two fields interfere constructively. The results corresponding to the 
parameters of Fig. \ref{ion_phase} with $I_f^{\text{max}} = 8\times 10^{10}$ 
W/cm$^2$ are collected in Figs. \ref{ioenph_z} and \ref{pulse_h}. One can 
see in Fig. \ref{ioenph_z} that, in the course of propagation, the relative 
phase $\theta$ (taken at the dynamic pulse maximum $t_{\text{max}}+z/c$) 
grows rapidly and over a distance of the order of 1 cm reaches the 
value $\pi$, at which the initial constructive interference of the two fields 
turns to destructive. At the same time, the total energy of the harmonic 
pulse, after a small reduction over a short interval of $z$, begins 
to increase as a result of the energy transfer from the strong fundamental 
field, in the parametric conversion process. This small reduction of the 
harmonic takes place only at the beginning of the propagation, when the 
relative phase is still close to 0 and the two fields interfere 
constructively, in the process of excitation of atoms from the ground state 
$\ket{1}$ to the state $\ket{2}$, while the generated part of the harmonic 
field is out of phase with the fundamental approximately by $\pi$ and 
continues to build up with a slight oscillation around the value $\pi$ of the
phase. It is important to mention that throughout the propagation, the 
amplitude and the phase of the fundamental field do not change significantly. 
This is because the number of photons contained in that pulse exceeds by 
many ($\geq 6$) orders of magnitude the number of atoms the pulse interacts 
with over the distance of $z \leq 20$ cm. Comparing the three graphs of 
Fig. \ref{ioenph_z}, one can see that with increasing of $\theta$ and $S_h$, 
the ionization probability first also grows, which is consistent with the 
previous discussion related to that intensity of the fundamental filed. But 
as $\theta$ approaches $\pi$, the ion yield drops almost exponentially until 
$Q \simeq 10 \%$. This residual ionization that is present even at 
$\theta \simeq \pi$ (and tends to 0 rather slowly) is caused by the fact 
that, because of the significant increase of the total energy of harmonic 
field, conditions (\ref{ampl_wid}) are not completely satisfied and 
the upper atomic level $\ket{2}$ acquires population due to that fraction 
of the generated field which exceeds the initial. Since the temporal widths 
of the pulses are less than the (radiative) relaxation time of the atomic 
coherence $\sigma_{21}$ ($\ga^{-1}\simeq 2$ ns), a significant fraction of 
the harmonic pulse amplitude is generated behind the fundamental 
(Fig. \ref{pulse_h}). That part of the amplitude is then attenuated due to the
atomic relaxation. Thus the total energy of the harmonic, after passing a
maximum at $z \simeq 5-7$ cm, decays then slowly back. Under these conditions,
the leading part of the harmonic pulse that falls under the temporal shape 
of the fundamental is by $\theta \simeq \pi$ out of phase with the latter 
and therefore the ionization vanishes, while the generated tail is 
continuously scattered by the atoms in the process of radiative decay.
The oscillations of the relative phase around $\pi$ are also slowly damped 
and the propagation reaches a ``dynamic equilibrium''

We note finally that a similar behavior of the system is obtained for a range 
of intensities we have explored. The main difference is that for weaker fields
($I_f^{\text{max}} = 3\times 10^{10}$ and $1\times 10^{10}$ W/cm$^2$) the ion 
yield does not exhibit a maximum other than at $z=0$ and drops to zero much 
faster as $z$ increases, which is consistent with the discussion above.

We have examined the problem of propagation in the simplest context of phase 
control, namely the excitation of a bound state, which has been used as a 
prototype in much of the initial work \cite{shapiro,elliott}. As discussed
here, the problem bears resemblance to earlier works 
\cite{co_mi,ja_wy,char,elt} on cancellation in third harmonic generation 
experiments, and so does the whole issue of phase control. The relevance 
and possible impact of propagation has been recognized by Chen and Elliott
\cite{elliott_pr} who presented data and an interpretation in terms of rate 
equations \cite{elt}.  Their study showed evidence of non-linear coupling, 
such as those discussed above, and called for ``more rigorous techniques'' in 
the approach to this basic problem. In the limit of validity of rate equations,
our results do indeed recapture the equations employed in their analysis.  
It will be interesting to explore this issue under more general conditions, 
such as the excitation of states embedded in continua, on which we expect to 
report elsewhere. The basic features of our analysis should, however, remain 
valid.

In summary, we have shown that the propagation of a bichromatic field with a 
preselected initial relative phase, has a profound effect. Over a rather
short scaled distance and independent of its initial value, the relative 
phase settles to a value that makes the medium transparent to the radiation, 
precluding thus further excitation and consequently control. The scaled 
distance $z N \Sigma$ does of course involve the density of the species and 
the cross-section of the laser beam, which suggests some flexibility on the 
choice of these parameters. In any case, however, the actual length of the 
interaction region over which control can be active will be defined and 
limited by the combination of the above parameters, as well as by the geometry
of the focused or unfocused laser beam. Briefly, for not very low atomic 
densities ($N>10^{12}$ cm$^{-3}$), the harmonic field settles to the 
steady-state value within a thin layer where a focused beam is well 
approximated by a plane wave. In the presence of large ac Stark shifts, 
however, a detailed analysis including specific experimental parameters is 
mandatory.

\begin{references}

\bibitem{shapiro} M. Shapiro, J. W. Hepburn, and P. Brumer,
Chem. Phys. Lett. {\bf 149}, 451 (1988);
C. K. Chan, P. Brumer, and M. Shapiro,
J. Chem. Phys. {\bf 94}, 2688 (1991).

\bibitem{elliott} C. Chen, Y. Yin, and D. S. Elliott,
Phys. Rev. Lett. {\bf 64}, 507 (1990);
C. Chen, and D. S. Elliott,
{\it ibid.} {\bf 65}, 1737 (1990).

\bibitem{zhu} L. Zhu, K. Suto, J. A. Fiss, R. Wada, 
T. Seideman, and R. J. Gordon, 
Phys. Rev. Lett. {\bf 79}, 4108 (1997);
L. Zhu, V. Kleiman, X. Li, S. Lu, K. Trentelman, and R. J. Gordon, 
Science, {\bf 270}, 77 (1995) and references therein.

\bibitem{nl} T. Nakajima and P. Lambropoulos, 
Phys. Rev. Lett. {\bf 70}, 1081 (1993); 
P. Lambropoulos and T. Nakajima, 
{\it ibid.} {\bf 82}, 2266 (1999); 
T. Nakajima and P. Lambropoulos, 
Phys. Rev. A {\bf 50}, 595 (1994); 
T. Nakajima, J. Zhang, and P. Lambropoulos,
J. Phys. B {\bf 30}, 1077 (1997).

\bibitem{cl}  J. C. Camparo and P. Lambropoulos, 
Phys. Rev. A {\bf 55}, 552 (1997); {\bf 59}, 2515 (1999).

\bibitem{elliott_pr} Ce Chen and D. S. Elliott,
Phys. Rev. A {\bf 53}, 272 (1996).

\bibitem{co_mi} R. N. Compton, J. C. Miller, A. E. Carter, and P. Kruit,
Chem. Phys. Lett. {\bf 71}, 87 (1980);
J. C. Miller, R. N. Compton, M. G. Payne, and W. W. Garret,
Phys. Rev. Lett. {\bf 45}, 114 (1980).

\bibitem{ja_wy} D. J. Jackson and J. J. Wynne,
Phys. Rev. Lett. {\bf 49}, 543 (1982);
J. J. Wynne,
{\it ibid.} {\bf 52}, 751 (1984);
D. J. Jackson, J. J. Wynne, and P. H. Kes,
Phys. Rev. A {\bf 28}, 781 (1983).

\bibitem{char} D. Charalambidis, X. Xing, J. Petrakis, and C. Fotakis,
Phys. Rev. A {\bf 44}, R24 (1991).

\bibitem{elt} M. Elk, P. Lambropoulos, and X. Tang,
Phys. Rev. A {\bf 46}, 465 (1992). 

\end{references}
\end{document}